\documentclass[12pt]{article}
\usepackage{mathrsfs}

\usepackage{fullpage}
\usepackage{amsfonts}
\usepackage{graphicx}
\usepackage{mathrsfs}
\usepackage{amsmath}
\usepackage{amssymb}
\usepackage{float}

\newtheorem{thm}{Theorem}
\newtheorem{lemma}{Lemma}
\newtheorem{coro}{Corollary}

\def\proof {{\noindent\bf Proof.}\quad}
\def\var{\mathrm {var}}

\newcommand{\bm}{\boldsymbol}

\def\tr{\mathrm {tr}}
\def\U{{\bm U}}
\def\A{{\bf A}}
\def\M{{\bf M}}

\def\D{{\bf D}}
\def\R{{\bm R}}

\def\S{{\bf S}}
\def\I{{\bf I}}

\def\u{{\bm u}}
\def\v{{\bm v}}

\def\X{{\bm X}}

\def\V{\bm V}

\def\Y{{\bm Y}}
\def\BL{{\bf \Lambda}}

\def\tr{\mathrm {tr}}
\def\Z{{\bm Z}}
\def\bfg{{\bf \Gamma}}
\def\bmv{{\bm\varepsilon}}
\def\bmw{{\bm \omega}}
\def\bth{{\bm\theta}}
\def\bms{{\bf\Sigma}}
\def\bfo{{\bf \Omega}}
\def\bfx{{\bf \Xi}}
\def\diag{\mathrm {diag}}
\newcommand{\con}{\rightarrow}

\def\cd{\overset{\mathcal{L}}{\longrightarrow}}

\title{\bf High Dimensional Rank Tests for Sphericity}
\author{Long Feng\\
{\em \small Northeast Normal University}}
\date{}
\begin{document}
\maketitle
\begin{abstract}
Sphericity test plays a key role in many statistical problems. We
propose Spearman's rho-type rank test and Kendall's tau-type rank
test for sphericity in the high dimensional settings. We show that
these two tests are equivalent. Thanks to the ``blessing of
dimension", we do not need to estimate any nuisance parameters.
Without estimating the location parameter, we can allow the
dimension to be arbitrary large. Asymptotic normality of these two
tests are also established under elliptical distributions.
Simulations demonstrate that they are very robust and efficient in a
wide range of settings.

\vspace{0.2cm} {\bf Key words}: Asymptotic normality; Kendall's
tau-type rank test; Large $p$, small $n$; Spatial rank; Spatial
sign; Spearman's rho-type test;  Sphericity test.
\end{abstract}

\section{Introduction}

Let $\X_1,\ldots,\X_n$ be a random sample from a $p$-variate
elliptical random vectors with scatter matrix $\bms_p$, which
describes the covariances between the $p$ variables. We wish to test
the following hypothesis
\begin{align}\label{test}
H_0: \bms_p=\sigma \I_p ~~\text{v.s.}~~ H_1:\bms_p\not=\sigma \I_p.
\end{align}
Such test play a key role in a number of statistical problems. It aries from several areas of
statistical applications, such as microarray analysis, geostatistics. When the dimension $p$ is
fixed, there are a considerable body of literature on this sphericity testing problem. For multinormal variables, a
classical method to deal with this problem is the likelihood ratio test (Mauchly 1940). John (1971, 1972) proposed the statistic
\begin{align*}\label{john}
Q_{\mathrm{J}}=\frac{np^2}{2}\mbox{tr}\left\{\frac{\S}{\mbox{tr}(\S)}-\frac{1}{p}
\I_p\right\}^2
\end{align*}
where $\S$ is the sample covariance matrix. He show that it is
locally powerful invariant test for sphericity under the
multivariate normal assumption. Muirhead and Waternaux  (1980)
modified John's test statistic to a wider elliptical distribution.

With the rapid development of technology, various types of
high-dimensional data have been generated in many areas, such as
hyperspectral imagery, internet portals, microarray analysis and
DNA. In genomic studies the data dimension can be a lot larger than
the sample size, say a so-called ``large $p$, small $n$" case.
Recently, many efforts have been devoted to sphericity test in high
dimensional settings. Bai et al. (2009) propose a corrections to the
likelihood ratio test by random matrix theory when $p/n \to c \in
(0,1)$. Ledoit and Wolf (2002) show that the existing $n$-asymptotic
theory remains valid if $p$ goes to infinity with $n$, even for the
case $p>n$. Without the normal distribution assumption, Chen, Zhang
and Zhong (2010) proposed a high-dimensional test based on $Q_J$
with two accurate estimators for $\tr(\bms_p)$ and $\tr(\bms_p^2)$.
Without specifying explicitly growth rate of $p$ relative to $n$,
they showed that their proposed test statistic is asymptotically
normal under the diverging factor model (Bai and Saranadasa 1996).
Though the diverging factor model contains a wide range of
distributions, it is difficult to justify. Moreover, the
multivariate $t$-distribution or mixture of multivariate
distribution does not satisfy this model. This motivates us to
construct more robust tests for sphericity.

In the traditional fixed $p$ circumstance, multivariate sign- and/or
rank-based covariance matrices are often used to construct robust
test for sphericity. See Hallin and Paindaveine (2006) and Oja
(2010) for nice overviews of this topic. However, when the dimension
is lager than the sample sizes, these methods may not work very
well. Zou et al. (2014) showed that the type I error of those tests
based on multivariate signs, such as Marden and Gao (2002), Hallin
and Paindaveine (2006) and Sirki\"{a} et al. (2009), are much larger
than the nominal level because of the estimation of location
parameters. Thus, Zou et al. (2014) propose a bias correction
procedure to the existing test statistic. However, it only can allow
the dimension at most being the square of the sample sizes. In
practice, the dimension of microarray data may be the exponential
rate of the sample sizes. It motivates us to construct new tests for
this ultra-high dimensional cases.

When $p$ is fixed, Spearman's rho-type test and Kendall's tau-type
rank test are the other two robust and efficient tests for
sphericity (Sirki\"{a} et al. 2009). However, there are many
nuisance parameters in these procedures. And those estimators
proposed in Sirki\"{a} et al. (2009) are unrealistic for high
dimensional data because of complex calculation or the assumption of
original location. Moreover, those nature estimators of $\tr({\bf
\Omega}_p^2)$ or $\tr({\bf \Xi}^2_p)$ based on the sample
symmetrized sign or rank covariance matrix would result in a
non-negligible bias term when the dimension is ultra-high. In this
article, we propose two novel Spearman's rho-type test and Kendall's
tau-type rank test for sphericity in the high dimensional settings.
Thanks to the ``blessing of dimension", those parameters do not need
to estimate anymore. Based on the leave out method, there are no
bias term in out test statistics. Additionally, without estimating
the location parameter, we can allow the dimension to be arbitrary
large. Asymptotic normality of these two tests are also established
under elliptical distributions. Simulations also demonstrate that
the proposed methods work reasonably well not only for those
elliptical distribution but also for the diverging factor model.

\section{High-dimensional rank tests}
\subsection{High-dimensional Spearman's rho-type rank test
statistic}

Suppose $\X_1,\ldots,\X_n$ are generated from a $p$-variate
elliptical distribution with density function
$\mbox{det}(\bms_p)^{-1/2}g_p\{||\bms_p^{-1/2}(\X-\bth_p)||\}$,
where $||\X||=(\X^T\X)^{1/2}$ is the Euclidean length of the vector
$\X$, $\bth_p$ is the symmetry center and $\bms_p$ is a positive
definite symmetric $p\times p$ scatter matrix. Similar to Zou et al.
(2014), define $\bms_p=\sigma_p\BL_p$ where $\tr(\BL_p)=p$ and
$\sigma_p$ is a scaled parameter. The hypothesis test (\ref{test})
is equivalent to test
\begin{align*}
H_0: \BL_p=\I_p, ~~\text{vs}~~H_1:\BL_p\not=\I_p.
\end{align*}
The spatial-rank function is defined as $R(\X)=E(U(\X-\Y)|\X)$,
where $U(\X)=||\X||^{-1}\X I(\X\neq 0)$. The spatial-rank covariance
matrix is $\bfo_p=E(R(\X)R(\X)^T)$. Under the null hypothesis,
$\bfo_p=\tau_Fp^{-1}\I_p$ where $\tau_F$ is a constant dependent on
$g_p$. Similar to the John's test, a nature distance measure between
$\bfo_p$ and $\tau_Fp^{-1}\I_p$ is
\begin{align*}
p\tr\left(\frac{\bfo_{p}}{\tr(\bfo_{p})}-p^{-1}\I_p\right)^2=\frac{p\tr(\bfo_{p}^2)}{\tr^2(\bfo_{p})}-1.
\end{align*}
In the fixed $p$ cases, we adopt the sample spatial-rank covariance
matrix $\bfo_{n,p}$ to estimate $\bfo_p$, i.e.
\begin{align*}
\bfo_{n,p}=\frac{1}{n}\sum_{i=1}^n
\R_i\R_i^T=\frac{1}{n^3}\sum_{i=1}^n\sum_{j=1}^n\sum_{k=1}^n
\U_{ij}\U_{ik}^T
\end{align*}
where $\R_i=\frac{1}{n}\sum_{j=1}^n \U_{ij}$,
$\U_{ij}=U(\X_i-\X_j)$. Then, the Spearman's rho-type rank test
statistic is defined as
\begin{align*}
Q_S=p\tr\left(\frac{\bfo_{n,p}}{\tr(\bfo_{n,p})}-p^{-1}\I_p\right)^2=\frac{p\tr(\bfo_{n,p}^2)}{\tr^2(\bfo_{n,p})}-1
\end{align*}
It can be shown that when $p$ is fixed, under the null hypothesis one has
\begin{align*}
\frac{n}{\gamma_S/\tau_F^2}Q_S\cd \chi^2_{(p+2)(p-1)/2}
\end{align*}
where $\gamma_S$, $\tau_F$ are two nuisance parameters dependent on
$g_p$ and $p$. Sirki\"{a} et al. (2009) suggest that we can estimate
$\tau_F$ by $\tr(\bfo_{n,p})/p$. And they suggest two estimators for
$\gamma_S$. One is estimated from the defining formula of
$\gamma_S$. However, it must assume the location of $\X_i$ to be the
origin, which is unrealistic in practice. Additionally, if we
standardize the samples by the estimated location parameters, as
shown in Zou et al. (2014), there would be another non-negligible
bias term in $Q_S$ when $p/n^2$ is large enough. The other estimator
of $\gamma_S$ is a complex symmetric U-statistic, which requires
$O(n^5p^4)$ computation. And the total calculation of $Q_S$ is of
order $O(n^5p^4)+O(p^6)$ because of the inverse of covariance matrix
of $vec(\bfo_{n,p})$. It is a too complicated calculation for high
dimensional data.

Fortunately, according to Lemma 1 in the appendix,
$E(\bfo_p)=0.5p^{-1}\I_p(1+o(1))$ under the null hypothesis as $p\to
\infty$. Thus, $\tr(\bfo_p)\to 0.5$. Thus, we only need to propose a
better estimator for $\tr(\bfo_p^2)$. However, the nature estimator
$\tr(\bfo_{n,p}^2)$ would result in a non-negligible bias term in
$Q_S$ when $p$ is ultra-high. Based on the leave out method, we
define the following new estimator for $\tr(\bfo_p^2)$,
\begin{align*}
\widehat{\tr(\bfo_p^2)}=\frac{1}{2n(n-1)(n-2)(n-3)}\underset{i,j,k,l~are~not~equal}{\sum\sum\sum\sum}\U_{ij}^T\U_{kl}\U_{kj}^T\U_{il}
\end{align*}
 Then,
we define the following high dimensional Spearman's rho-type rank
test statistic (abbreviated as SR hereafter)
\begin{align*}
\tilde{Q}_S=4p\widehat{\tr(\bfo_p^2)}-1
\end{align*}
Obviously, the value of $\tilde{Q}_S$ remains unchanged for $\bm
Z_i=a {\bf O} \X_i+\bm c$ where $a$ is a constant, ${\bf O}$ is an
orthogonal matrix and $\bm c$ is a vector of constants. Thus, the
test statistic $\tilde{Q}_S$ is invariant under rotations. The
following theorem establishes the asymptotic null distribution of
$\tilde{Q}_S$.
\begin{thm}
Under $H_0$, as $n\to \infty$ and $p\to\infty$, $\tilde{Q}_S/\sigma_0 \cd N(0,1),$ where  $\sigma_0^2=4(p-1)/(n(n-1)(p+2))$.
\end{thm}
According to Theorem 1, there are not nuisance parameters in the new
proposed test procedure. As $n,p$ goes to infinity, $\tilde{Q}_S$ is
asymptotic normal and the variance is only dependent on $p$ and $n$.
It can be viewed as the phenomenon of ``blessing of dimension".
Moreover, the complexity of the entire procedure is only $O(n^4p)$,
which is eventually less than the classic Spearman's rho-type rank
test procedure.

Theorem 1 also shows that there is no bias term in $\tilde{Q}_S$.
So, we do not need a bias-correction procedure as Zou et al. (2014).
Moreover, we do not require the relationship between the sample size
$n$ and dimension $p$. However, the test proposed by Zou et al
(2014) (abbreviated as SS hereafter) must require the dimension
being the square of the sample size at most. When $p/n^2 \to
\infty$, there would be another bias-term in SS test statistic,
which is difficult to calculate. Simulation studies also demonstrate
these results. See more information in Section 3.

Next, we consider the asymptotic distribution of $\tilde{Q}_S$ under
the alternative $H_1: \BL_p=\I_p+\D_{n,p}$. Define
\[
{\sigma}_1^2={\sigma}_0^2+n^{-2}p^{-2}\left\{8p\tr(\D_{n,p}^2)+4\tr^2(\D_{n,p}^2)\right\}+8n^{-1}p^{-2}\left\{\tr(\BL_p^4)-p^{-1}\tr^2(\BL_p^2)\right\}.
\]

\begin{thm}
Suppose that $n\tr(\D_{n,p}^2)/p=O(1)$. Under $H_1$,
$\{\tilde{Q}_S-\tr(\D_{n,p}^2)/p\}/{{\sigma}_1}\cd N(0,1)$, as
$p\con\infty, n\con\infty$.
\end{thm}
According to Theorem 2, if $p=O(n^2)$, $\tilde{Q}_S$ has the same
power function as the test proposed by Zou et al. (2014). However,
when $p/n^2 \to \infty$, the variance of SS test statistic will be
larger than $\sigma_1^2$ because of the estimation of location
parameter $\bth_p$. See more discussion about it in Section 3.

In addition, we could establish the consistency of our
high-dimensional Spearman's rho-type rank test based on Theorem 2.
\begin{coro}
If $n\tr(\D_{n,p}^2)/p\to \infty$, the test
$\tilde{Q}_S/\sigma_0>z_{\alpha}$ is consistent against $H_1$ as
$n\to \infty$ and $p\to \infty$.
\end{coro}

Theorems 1 and 2 also allow us to compare our SR test with the existing work, such as Chen et al. (2010).
The following corollary concerns the limiting efficiency comparison between Chen et al. (2010) test
(abbreviated as CZZ hereafter) under multivariate normality assumption.
\begin{coro}
If $C_1< n\tr(\D_{n,p}^2)/p<C_2$, under multi-normal distributions,
SR test is asymptotically efficient as CZZ test.
\end{coro}
It is worth pointing out that theoretically comparing the proposed
test with CZZ test under general multivariate distributions turns
out to be difficult. This is because the asymptotic validity of CZZ
test relies on the diverging factor model, while elliptical
assumption is required in Theorems 1 and 2. The distinction and
connection between the elliptical distributions and the diverging
factor model is far from clear in the literature.

\subsection{High-dimensional Kendall's tau-type rank test statistic}

In this subsection, we consider another efficient sphericity test,
Kendall's tau-type rank test. The classic Keandal's tau covariance
matrix is defined as
$\bfx_{n,p}=\frac{2}{n(n-1)}\sum_{i<j}\U_{ij}\U_{ij}^T$. Under
$H_0$, we have $E(\bfx_{n,p})\doteq \bfx_p=p^{-1}\I_p$. Thus, the
Kendall's tau test statistic is defined as
\begin{align*}
Q_K=p\tr(\tr^{-1}(\bfx_{n,p})\bfx_{n,p}-p^{-1}\I_p)^2=p\tr(\bfx_{n,p}^2)-1
\end{align*}
It can be shown that when $p$ is fixed, under the null hypothesis one has
\begin{align*}
\frac{n}{\gamma_K}Q_K\cd \chi^2_{(p+2)(p-1)/2}
\end{align*}
where $\gamma_K$ is another nuisance parameter dependent on $g_p$
and $p$. Similarly, the estimator for $\gamma_K$ in Sirki\"{a} et
al. (2009) can not be used in high dimensional settings, which
requires original location or $O(n^3p^4)$ computation. Thanks for
the ``blessing of dimension", we also do not need this nuisance
parameter in high dimensional data. Moreover, the nature estimator
$\tr(\bfx_{n,p}^2)$ also would result in a non-negligible bias term
in $Q_K$ when $p$ is ultra-high. Thus, based on the leave out
method, we propose the following estimator for $\tr(\bfx_p^2)$,
\begin{align*}
\widehat{\tr(\bfx_p^2)}=\frac{1}{n(n-1)(n-2)(n-3)}\underset{i,j,k,l~are~not~equal}{\sum\sum\sum\sum}
(\U_{ij}^T\U_{kl})^2
\end{align*}
Then, we define the following high-dimensional Kendall's tau-type
rank test statistic (abbreviated as SK hereafter)
\begin{align*}
\tilde{Q}_K=p\widehat{\tr(\bfx_p^2)}-1
\end{align*}
Obviously, the test statistic $\tilde{Q}_K$ is also invariant under
rotations. We can also establish the asymptotic properties of
$\tilde{Q}_K$ as follow.
\begin{thm}
As $n\to \infty$ and $p\to \infty$,
\begin{itemize}
\item[(i)] Under $H_0$, $\tilde{Q}_K/\sigma_0 \cd N(0,1)$.
\item[(ii)] Under $H_1$, if $n\tr(\D_{n,p}^2)/p=O(1)$, $\{\tilde{Q}_K-\tr(\D_{n,p}^2)/p\}/{{\sigma}_1}\cd
N(0,1)$.
\end{itemize}
\end{thm}
In fact, as shown in the proof of Theorem 3, $\tilde{Q}_K$ is
asymptotic equivalent to $\tilde{Q}_S$ under both null and
alternative hypothesis. In high dimensional settings, the Kedall's
tau-type rank test is equivalent to the Spearman's rho-type rank
test. Thus, similar to Corollary 1, we can also show the consistency
of SK test. And SK test is also asymptotic efficient as CZZ test
under the multinormal distributions by the similar arguments as
Corollary 2. We state these results in the following corollary.
\begin{coro}
As $n\to \infty$ and $p\to \infty$, we have
\begin{itemize}
\item[(i)] if $n\tr(\D_{n,p}^2)/p\to \infty$, the test $\tilde{Q}_K/\sigma_0>z_{\alpha}$ is consistent against $H_1$.
\item[(ii)] if $C_1< n\tr(\D_{n,p}^2)/p<C_2$, under multi-normal distributions, SK test is asymptotically efficient as CZZ test.
\end{itemize}
\end{coro}

\section{Simulation}

We consider the following five distributions for comparison:
\begin{itemize}
\item[(I)] The standard multivariate normal;
\item[(II)]  The standard multivariate $t$ with four degrees of freedom, $t_{p,4}$;
\item[(III)] Mixtures of two multivariate normal densities $\kappa
f_p(\mu,\I_p)+(1-\kappa)f_p(\mu,9\I_p)$, where $f_p(\cdot;\cdot)$ is
the $p$-variate multivariate normal density. The value $\kappa$ is
chosen to be 0.8.
\item[(IV)] The diverging factor model with the standardized Gamma(4, 0.5) distribution;
\item[(V)] The diverging factor model with the standardized $t$ distribution with four degrees of freedom, $t_{4}$.
\end{itemize}
Here we choose $\bfg=\I_p$ and for each $\Z_i$, $p$ independent
identically distributed random variables $Z_{ij}$'s are generated in
diverging factor model in Scenarios (IV) and (V). The first three
scenarios are the well-known multivariate elliptical distributions.
However, the last two scenarios are not elliptically distributed. We
consider the sample sizes $n=20,30$ and dimensions
$p=100,200,400,800$. Similar to Chen et al. (2010), we obtain the
observations $\X_i=\A\Y_i$, where $\Y_i$ are generated from Scenario
(I)-(V) and $\A=\diag\{2^{1/2}1_{[vp]}, 1_{p-[vp]}\}$, $[x]$ denotes
the integer truncation of $x$. Three levels of $v$ were considered:
0(size), 0.15 and 0.3. We compare our high-dimensional Spearman's
rho-type rank test (abbreviated as SR), high-dimensional Kendall's
tau test (abbreviated as SK) with the bias-corrected sign test
proposed by Zou et al. (2014) (abbreviated as SS) and the sphericity
test proposed by Chen et al. (2010)(abbreviated as CZZ). Tables
\ref{t1} and \ref{t2} report the empirical sizes and power of these
four tests under Scenarios (I)-(III), (IV)-(V), respectively.

\begin{figure}
\begin{center}
\includegraphics[width=14.0cm,height=8cm]{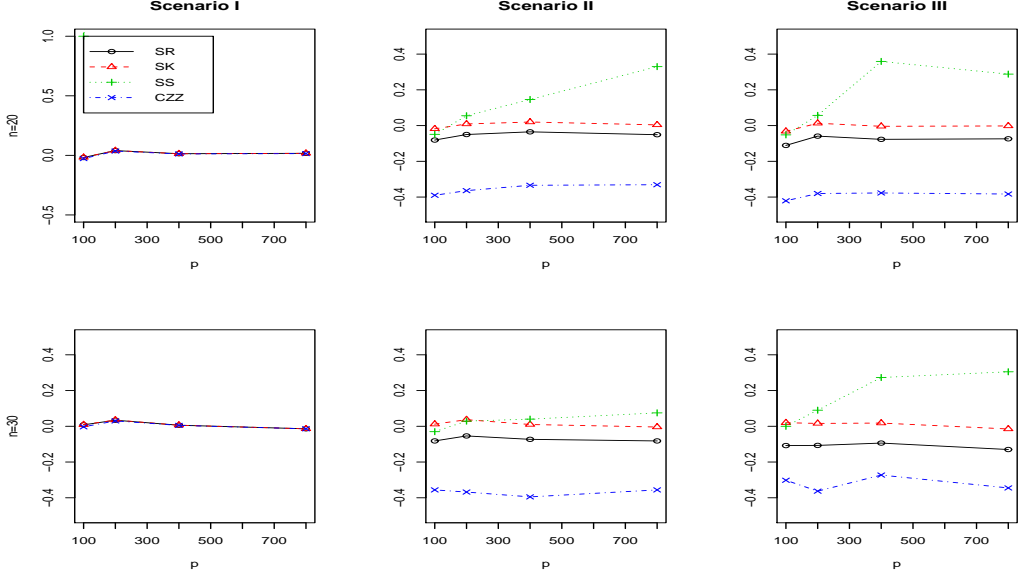}\vspace{-0.3cm}
\end{center}
\caption{The mean-standard deviation-ratio of test statistics under Scenarios (I)-(III).}\label{figure1}\vspace{-0.2cm}
\end{figure}

\begin{figure}
\begin{center}
\includegraphics[width=12.0cm,height=8cm]{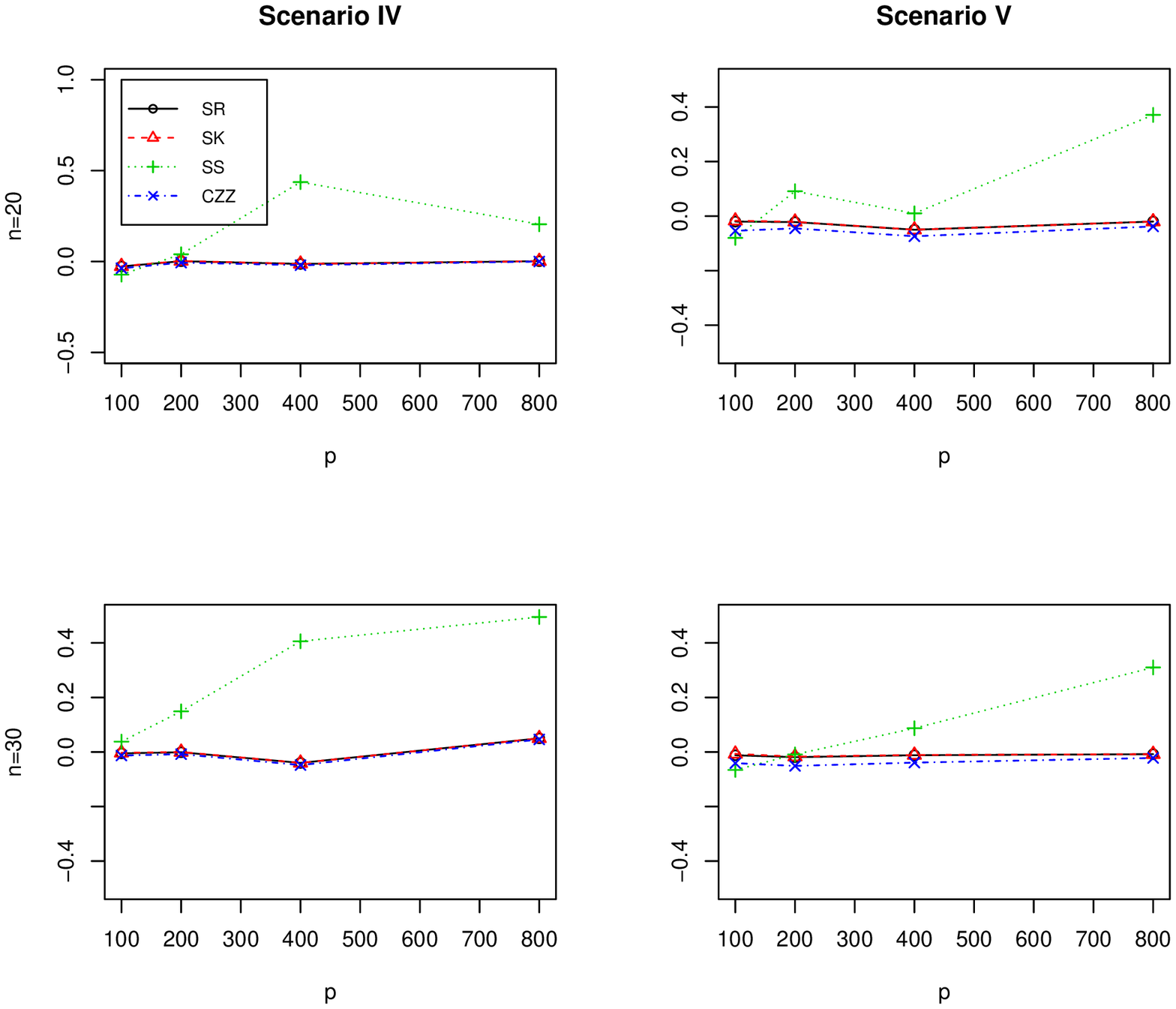}\vspace{-0.3cm}
\end{center}
\caption{The mean-standard deviation-ratio of test statistics under Scenarios (IV)-(V).}\label{figure2}\vspace{-0.2cm}
\end{figure}

\begin{figure}
\begin{center}
\includegraphics[width=14.0cm,height=8cm]{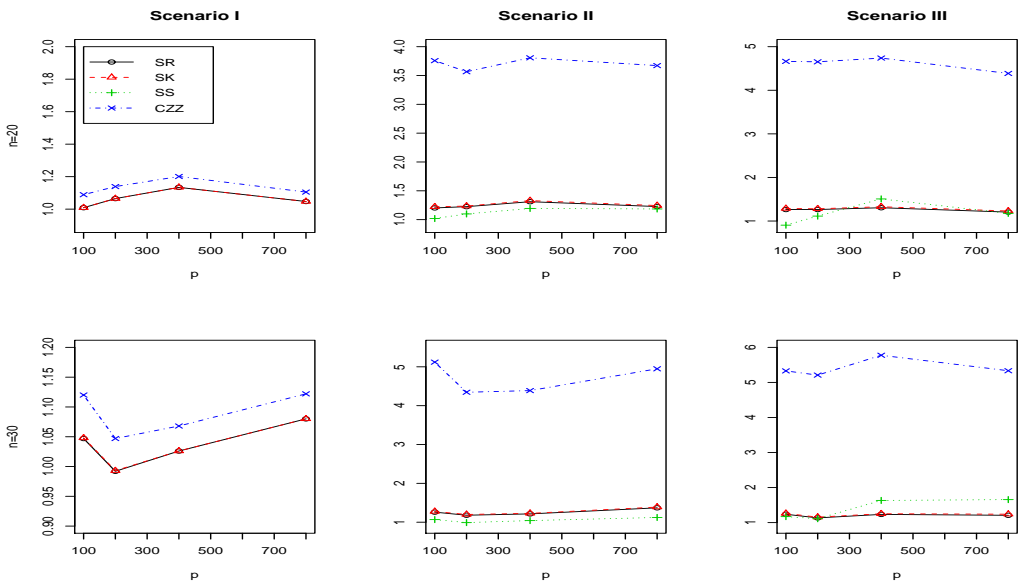}\vspace{-0.3cm}
\end{center}
\caption{The variance-ratio of test statistics under Scenarios (I)-(III).}\label{figure3}\vspace{-0.2cm}
\end{figure}

\begin{figure}
\begin{center}
\includegraphics[width=12.0cm,height=8cm]{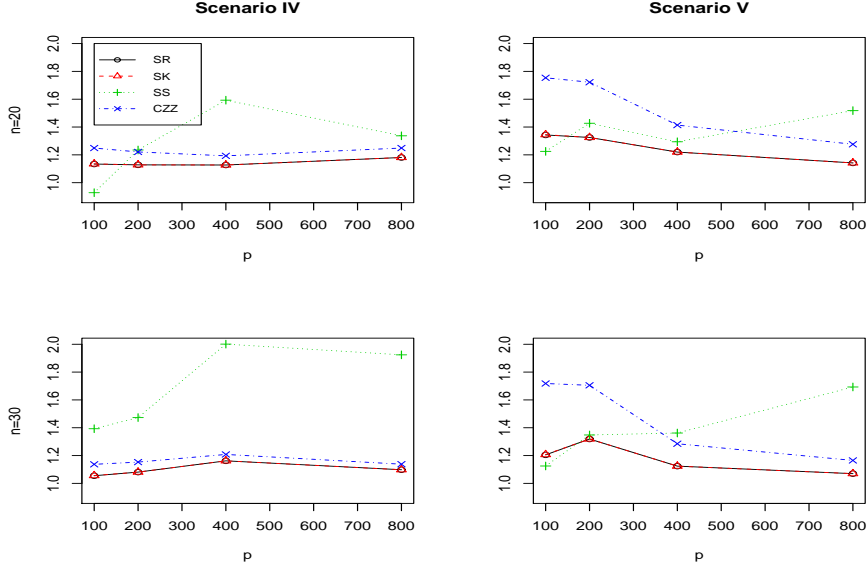}\vspace{-0.3cm}
\end{center}
\caption{The variance-ratio of tests under Scenarios (IV)-(V).}\label{figure4}\vspace{-0.2cm}
\end{figure}

Firstly, we consider the empirical sizes of these tests. The
empirical sizes of SR and SS tests are close to the nominal level in
al cases, which is not impacted by the dimension. However, SS can
not control its empirical sizes very well in many cases. Sometimes
it is a little conservative but sometimes it is too larger than the
nominal level. To evaluate the impact of dimension to the bias-term
of SS, we also report the mean-standard deviation-ratio
$E(T)/\sqrt{\var(T)}$ and the variance estimator ratio
$\var(T)/\widehat{\var(T)}$ of these four tests. Since the explicit
form of $E(T)$ and $\var(T)$ is difficult to calculate for all
tests, we estimate them by simulation. Figures \ref{figure1} and
\ref{figure2} report the mean-standard deviation-ratio of these four
tests. Figures \ref{figure3} and \ref{figure4} report the variance
estimator ratio of these tests. We observe that the bias term in SS
is apparently exists, especially when $p/n^2$ is large. It is not
strange because SS can only allow the dimension being comparable to
the square of the sample size. In contrast, the mean-standard
deviation-ratio of our SR and SK test statistics is approximately
zero, which shows that, regardless of the dimension, there is no
bias-term in our test statistics. Under scenario (III)-(V), the
variance estimator ratio of SS is eventually larger than one when
$p/n^2$ is large. When the dimension gets larger, the bias of
spatial-median estimator will also increase the variance of SS test
statistic. So the empirical sizes of SS is difficult to maintain in
these cases. However,  the variance estimator ratio of our SR and SK
test statistic is approximately one. Without estimating the location
parameter, the variance of SR and SK test statistic do not increase
with the dimension. In addition, when the sample are generated from
the diverging factor model, the empirical sizes of CZZ test are a
little larger than the nominal level in most cases. However, under
Scenario (II) and (III), the mean-standard deviation-ratio of CZZ is
smaller than zero and the variance estimator ratio is eventually
larger than one. And then, the empirical sizes of CZZ test are
significantly larger than the nominal level. It is not surprising
because neither $t_{p,4}$ nor a mixture of multivariate normal
distributions belongs to the diverging factor model.

Next, we consider the power comparison of these tests.  SR and SK
tests perform similar to each other, which is consistent with the
theoretical results in section 2. In general, both SR and SK tests
perform a little better than SS test in most cases. The variance of
SS test statistic will increase faster than SR and SK test
statistics because of the estimation of location parameters. Then it
is not surprising that the power of SS is smaller than these two
tests. Moreover, the power of SS is larger than SR and SK in some
cases, such as scenario II with $(n,p)=(20,800)$. However, the
empirical sizes of SS also are lager than the nominal level in these
cases. Thus its high power would not be very meaningful. In
addition, our SR and SK test perform similar to CZZ test under
normal distributions. Even under the non-elliptical distributions
(Scenarios (IV) and (V)), the difference between CZZ and SR and SK
is marginal. However, under two heavy-tailed elliptical
distributions (Scenario (II) and (III)), our SR and SK tests
performs eventually better than CZZ test.

All these results suggest that the proposed two test are quite
robust and efficient in testing sphericity. Without estimating the
location parameter, SR and SK tests can control their empirical
sizes very well and are more powerful than SS test under the
alternative hypothesis. For heavy-tailed or skewed distributions, SR
and SK tests performs much better than CZZ test both in sizes and
power.

 \begin{table}
           \centering
           \caption{Empirical Size and  power comparison at 5\% significance under Scenarios (I)-(III)}
           \vspace{0.1cm}
      \renewcommand{\arraystretch}{1.2}
     \tabcolsep 6pt
         \begin{tabular}{ccccccccccccccc}\hline \hline
    &  \multicolumn{4}{c}{Size}  &   & \multicolumn{4}{c}{$v=0.15$}&  &  \multicolumn{4}{c}{$v=0.30$}     \\
          {$(n,p)$}                 &  \multicolumn{1}{c}{SR}  & \multicolumn{1}{c}{SK} & \multicolumn{1}{c}{SS}   & \multicolumn{1}{c}{CZZ}&    &  \multicolumn{1}{c}{SR}  & \multicolumn{1}{c}{SK} & \multicolumn{1}{c}{SS}   & \multicolumn{1}{c}{CZZ}&  &    \multicolumn{1}{c}{SR}  & \multicolumn{1}{c}{SK} & \multicolumn{1}{c}{SS}   & \multicolumn{1}{c}{CZZ}\\\hline
          \multicolumn{15}{c}{Scenario (I)} \\
(20,100)   &  5.8  &  5.8  &  3.9  &  5.8   && 24  & 24  & 16   & 26  && 33  & 33  & 25  & 34  \\
(20,200)   &  6.3  &  6.3  &  5.3  &  6.5   && 28  & 28  & 23   & 29  && 36  & 36  & 22  & 36  \\
(20,400)   &  6.3  &  6.3  &  4.5  &  7.6   && 26  & 26  & 14   & 27  && 34  & 33  & 20  & 35  \\
(20,800)   &  6.0  &  6.0  &  6.0  &  7.6   && 25  & 25  & 21   & 26  && 36  & 36  & 21  & 37  \\
(30,100)   &  5.6  &  5.7  &  5.2  &  6.1   && 39  & 39  & 34   & 41  && 52  & 52  & 48  & 55  \\
(30,200)   &  4.9  &  4.9  &  3.6  &  5.5   && 42  & 42  & 34   & 43  && 56  & 56  & 51  & 56  \\
(30,400)   &  5.1  &  5.1  &  3.0  &  5.1   && 40  & 40  & 22   & 41  && 56  & 56  & 43  & 57  \\
(30,800)   &  6.5  &  6.5  &  4.2  &  6.8   && 41  & 41  & 30   & 42  && 55  & 55  & 47  & 56  \\ \hline
                 \multicolumn{15}{c}{Scenario (II)} \\
 (20,100)  &  5.0  &  5.3  &  5.8  &  9.7   && 24  & 26  & 23  & 21   && 30  & 32  & 32  & 25  \\
 (20,200)  &  4.9  &  5.8  &  6.8  & 10.1   && 26  & 28  & 28  & 22   && 32  & 35  & 35  & 27  \\
 (20,400)  &  5.9  &  6.7  &  9.0  & 11.5   && 25  & 27  & 28  & 22   && 32  & 34  & 34  & 27  \\
 (20,800)  &  5.0  &  5.7  & 11.7  & 10.1   && 24  & 26  & 33  & 22   && 34  & 37  & 45  & 28  \\
 (30,100)  &  5.7  &  4.9  &  5.3  & 11.6   && 37  & 40  & 38  & 28   && 48  & 51  & 50  & 34  \\
 (30,200)  &  6.0  &  5.6  &  5.5  & 11.0   && 40  & 43  & 41  & 30   && 52  & 56  & 55  & 39  \\
 (30,400)  &  5.2  &  5.2  &  6.4  & 10.8   && 38  & 41  & 41  & 30   && 52  & 55  & 57  & 37  \\
 (30,800)  &  6.5  &  6.0  &  7.9  & 12.0   && 38  & 41  & 42  & 31   && 50  & 53  & 57  & 38  \\ \hline
                 \multicolumn{15}{c}{Scenario (III)} \\
 (20,100)  &  6.2  &  6.2  &  4.8  & 11.4   && 21  & 23  & 21  & 19   && 29  & 31  & 28  & 23  \\
 (20,200)  &  5.9  &  5.8  &  6.7  & 12.2   && 25  & 27  & 26  & 22   && 32  & 35  & 30  & 25  \\
 (20,400)  &  5.8  &  6.3  &  5.0  & 12.7   && 25  & 27  & 23  & 21   && 34  & 35  & 28  & 24  \\
 (20,800)  &  5.2  &  5.9  &  9.2  & 11.9   && 24  & 27  & 29  & 21   && 34  & 37  & 29  & 26  \\
 (30,100)  &  4.6  &  6.3  &  5.3  & 14.9   && 36  & 41  & 38  & 31   && 48  & 54  & 50  & 37  \\
 (30,200)  &  4.8  &  4.5  &  4.6  & 13.7   && 38  & 42  & 41  & 29   && 50  & 54  & 54  & 35  \\
 (30,400)  &  5.7  &  5.5  &  3.6  & 16.8   && 37  & 41  & 36  & 31   && 52  & 57  & 54  & 37  \\
 (30,800)  &  5.8  &  5.0  &  5.9  & 13.4   && 37  & 41  & 40  & 28   && 51  & 55  & 55  & 35  \\ \hline \hline
               \end{tabular}\label{t1}
           \end{table}

 \begin{table}[ht]
           \centering
           \caption{Empirical Size and  power comparison at 5\% significance under Scenarios (IV)-(V)}
           \vspace{0.1cm}
      \renewcommand{\arraystretch}{1.2}
     \tabcolsep 6pt
         \begin{tabular}{ccccccccccccccc}\hline \hline
    &  \multicolumn{4}{c}{Size}  &   & \multicolumn{4}{c}{$v=0.15$}&  &  \multicolumn{4}{c}{$v=0.30$}     \\
          {$(n,p)$}                 &  \multicolumn{1}{c}{SR}  & \multicolumn{1}{c}{SK} & \multicolumn{1}{c}{SS}   & \multicolumn{1}{c}{CZZ}&    &  \multicolumn{1}{c}{SR}  & \multicolumn{1}{c}{SK} & \multicolumn{1}{c}{SS}   & \multicolumn{1}{c}{CZZ}&  &    \multicolumn{1}{c}{SR}  & \multicolumn{1}{c}{SK} & \multicolumn{1}{c}{SS}   & \multicolumn{1}{c}{CZZ}\\\hline
          \multicolumn{15}{c}{Scenario (IV)} \\
(20,100)   &  4.8  &  5.9  &  4.9  &  7.1   && 24  & 24  & 18  & 25   && 31  & 31  & 25  & 32  \\
(20,200)   &  5.0  &  5.0  &  5.8  &  7.8   && 27  & 27  & 23  & 28   && 34  & 34  & 25  & 35  \\
(20,400)   &  4.5  &  4.5  &  3.4  &  7.0   && 26  & 26  & 15  & 27   && 33  & 33  & 20  & 34  \\
(20,800)   &  5.0  &  5.0  &  6.6  &  7.4   && 25  & 25  & 22  & 26   && 35  & 35  & 19  & 36  \\
(30,100)   &  4.8  &  4.8  &  4.6  &  6.0   && 38  & 38  & 35  & 42   && 51  & 51  & 49  & 53  \\
(30,200)   &  5.6  &  5.8  &  4.7  &  6.1   && 40  & 40  & 36  & 42   && 55  & 55  & 52  & 56  \\
(30,400)   &  5.3  &  5.3  &  4.2  &  5.7   && 41  & 41  & 29  & 40   && 55  & 55  & 41  & 56  \\
(30,800)   &  5.9  &  4.9  &  3.8  &  7.1   && 42  & 42  & 33  & 43   && 57  & 57  & 49  & 57\\ \hline
                 \multicolumn{15}{c}{Scenario (V)} \\
(20,100)   &  5.5  &  5.5  &  5.9  &  9.8   && 25  & 25  & 20  & 27   && 30  & 30  & 26  & 32  \\
(20,200)   &  4.9  &  5.9  &  5.8  &  9.7   && 27  & 27  & 18  & 28   && 35  & 35  & 26  & 35  \\
(20,400)   &  4.6  &  5.6  &  5.6  &  6.8   && 25  & 25  & 21  & 27   && 32  & 32  & 26  & 34  \\
(20,800)   &  5.7  &  5.7  &  4.9  &  7.6   && 27  & 27  & 19  & 28   && 36  & 36  & 26  & 37  \\
(30,100)   &  4.2  &  4.2  &  5.8  &  8.4   && 36  & 36  & 33  & 39   && 50  & 49  & 45  & 51  \\
(30,200)   &  5.9  &  5.9  &  6.2  &  8.3   && 37  & 37  & 33  & 38   && 50  & 50  & 44  & 49  \\
(30,400)   &  4.5  &  4.5  &  5.0  &  7.1   && 40  & 40  & 32  & 40   && 54  & 54  & 50  & 55  \\
(30,800)   &  4.1  &  5.1  &  4.7  &  7.1   && 40  & 40  & 32  & 41   && 55  & 55  & 47  & 55\\ \hline \hline
               \end{tabular}\label{t2}
           \end{table}

\section{Discussion}

Multivariate-rank based method is very robust and efficient in
constructing test procedure in multivariate problems. In this paper,
we proposed two novel test statistic for sphericity test based on
multivariate-rank. We believe that this procedure can be extended to
more general elliptical distributions with
$\bms_p=\diag\{\sigma_{11},\cdots,\sigma_{pp}\}$ where the
$\sigma_{ii}$ are unknown. Moreover, high dimensional location
testing problem also draw much attention in statistics (Chen and Qin
2010). Wang et al. (2015) proposed a high dimensional test for one
sample location problem based on multivariate-sign. However, the
tests for location problem based on multivariate-rank deserve future
study in high-dimensional settings.

\section{Appendix}

\vspace{0.2cm}
\noindent{\bf Appendix A: Some useful Lemmas}

Denote $\bmv_i=\bms_p^{-1/2}(\X_i-\bth_p)$ and
$\u_i=E(U(\bmv_i-\bmv_j)|\bmv_i)$. Obviously,
$E(\u_i\u_i^T)=\tau_Fp^{-1}\I_p$ where $\tau_F$ is a constant depend
on distribution $g_p$ and $p$.
\begin{lemma}
$\tau_F\to 0.5$ as $p\to \infty$.
\end{lemma}
\proof
\begin{align*}
E(\bmv_i^T\bmv_i)=&E((\bmv_i-\bmv_j)^T(\bmv_i-\bmv_k))\\
=&E(E((\bmv_i-\bmv_j)^T(\bmv_i-\bmv_k)\big|\bmv_i))\\
=&E(E(||\bmv_i-\bmv_j||||\bmv_i-\bmv_k||U(\bmv_i-\bmv_j)^TU(\bmv_i-\bmv_k)\big|\bmv_i))\\
=&E((E(||\bmv_i-\bmv_j||\big|\bmv_i))^2)E(E(U(\bmv_i-\bmv_j)^TU(\bmv_i-\bmv_k)\big|\bmv_i))\\
=&E((E(||\bmv_i-\bmv_j||\big|\bmv_i))^2)E(\u_i^Tu_i)=\tau_FE((E(||\bmv_i-\bmv_j||\big|\bmv_i))^2)
\end{align*}
In addition, $E(||\bmv_i||^2)=0.5E(||\bmv_i-\bmv_j||^2)$. Thus, we
only need to show that
\begin{align*}
\frac{E((E(||\bmv_i-\bmv_j||\big|\bmv_i))^2)}{E(||\bmv_i-\bmv_j||^2)}\to
1.
\end{align*}
Because $\bmv_i$ has the elliptical distribution, $\bmv_i-\bmv_j$
also has the elliptical distribution. Define the density function of
$||\bmv_i-\bmv_j||$ is $f(t)=c_pt^{p-1}g(t)$ where
$c_p=\frac{2\pi^{p/2}}{\Gamma(p/2)}$. Thus,
\begin{align*}
\frac{E((E(||\bmv_i-\bmv_j||\big|\bmv_i))^2)}{E(||\bmv_i-\bmv_j||^2)}=&\frac{\left(\int c_p t^p g(t) dt\right)^2}{\int c_p t^{p+1} g(t) dt}\\
=&\frac{c_{p+1}^2}{c_pc_{p+2}}=\frac{\Gamma^2((p+1)/2)}{\Gamma(p/2)\Gamma((p+2)/2)}
\end{align*}
By the Stirling's formula,
\[
\lim_{x \rightarrow \infty}\frac{\Gamma(x+1)}{(x/e)^{x}(2\pi x)^{1/2}}=1,\]
as $p\to \infty$, we have
\begin{align*}
\frac{c_{p+1}^2}{c_pc_{p+2}} \to \frac{(p-1)^{p-1}}{p^{p/2}(p-2)^{(p-2)/2}}=(1-p^{-1})^{p/2}(1+(p-2)^{-1})^{(p-2)/2} \to 1.
\end{align*}
Here we complete the proof. \hfill$\Box$

\begin{lemma}
For any matrix $\M$, we have $E(\u_{j}^{T}\M
\u_{j})^2=O\left(p^{-2}\tr(\M^{T}\M)\right)+O(p^{-2}\tr^2(\M)),j=1,\cdots,n.$
\end{lemma}
\proof  Define $\M =(a_{lk})_{l,k=1}^p$,
$\u_{i}=(u_{i1},\ldots,u_{ip})^{T}$, so
\begin{align*}
E((\u_{i}^{T}\M\u_{i})^2)=&E\left(\left(\sum_{l,k=1}^p
a_{lk}u_{il}u_{ik}\right)^2\right)=\sum_{l,k=1}^p\sum_{s,t=1}^p
a_{lk}a_{st}E(u_{il}u_{ik}u_{is}u_{it})\\
=&\sum_{k=1}^p\sum_{l=1}^p a_{kl}^2 E(u_{ik}^2u_{il}^2)+\sum_{k=1}^p\sum_{l=1}^p a_{ll}a_{kk} E(u_{ik}^2u_{il}^2)
\end{align*}
Because $E(u_{il}^4)=O(p^{-2})$,
$E(u_{il}^2u_{il}^2)=O(p^{-2})$ and
\begin{align*}
\sum_{k=1}^p\sum_{l=1}^p a_{kl}^2=\tr(\M^T\M),
\sum_{k=1}^p\sum_{l=1}^p a_{ll}a_{kk}=\tr^2(\M).
\end{align*}
Thus,
$E(\u_{i}^{T}\M\u_{i})^2=O\left(p^{-2}\tr(\M^{T}\M)\right)+O\left(p^{-2}\tr^2(\M)\right).$
\hfill$\Box$

\begin{lemma}
As $n\to \infty$ and $p\to \infty$,
\begin{align*}
\frac{\frac{p}{n(n-1)}\underset{i\not=j}{\sum\sum}(\u_i^T\u_j)^2/\tau_F^2-1}{\sigma_0}\cd
N(0,1)
\end{align*}
\end{lemma}

\proof Define $\v_i=\u_i/\sqrt{\tau_F}$. Thus,
$E(\v_i\v_i^T)=p^{-1}\I_p$. Define
\[Q_s^{'}=\frac{p}{n(n-1)}\underset{i\not=j}{\sum\sum}(\u_i^T\u_j)^2/\tau_F^2-1=\frac{p}{n(n-1)}\underset{i\not=j}{\sum\sum}(\v_i^T\v_j)^2-1\]
The expectation of $Q_{\mathrm{S}}'$ can be easily verified and thus
omitted here. $\var(Q_{\mathrm{S}}')$ can be computed as follows:
\begin{align*}
\var(Q_{\mathrm{S}}')=&\{n(n-1)\}^{-2}p^2 E\left\{\sum_{i \neq j} (\v_i^T\v_j)^2\right\}^2 - 1 \\
=&\{n(n-1)\}^{-2}p^2 \big [ 2n(n-1) E(\v_i^T\v_j)^4 + 4n(n-1)(n-2) E \left\{(\v_i^T\v_j)^2(\v_i^T\v_k)^2 \right\} \\
&+n(n-1)(n-2)(n-3)E\left\{(\v_i^T\v_j)^2(\v_k^T\v_l)^2 \right\} \big ] -1 \\
=&4(p-1)/\{n(n-1)(p+2)\}.
\end{align*}
Next, we only need to show the asymptotic normality of $Q_S^{'}$.
Let $\mathcal{F}_0=\{\varnothing,\bfo\}$,
$\mathcal{F}_k=\sigma\{\v_1,\ldots,\v_k\}, k = 1,\ldots,n$. Let
$E_k(\cdot)$ denote the conditional expectation of given
$\mathcal{F}_k$ and $E_0(\cdot)= E(\cdot)$. Write
$Q_{\mathrm{S}}'-E(Q_{\mathrm{S}}')=\sum_{k=1}^nG_{n,k}$, where
$G_{n,k}=(E_k-E_{k-1})Q_{\mathrm{S}}'$. Then for every $n$,
$\{G_{n,k}\}_{k=1}^n$ is a martingale difference sequence with
respect to the $\sigma$-fields $\{\mathcal{F}_k,1\leq k\leq n\}$.
Let $\sigma_{n,k}^2=E_{k-1}(G_{n,k}^2)$. According to the martingale
central limit theorem (Hall and Hyde 1980), we only need to show
that, as $n\con\infty$,
\begin{align}\label{lyacons}
\frac{\sum_{k=1}^n\sigma^2_{n,k}}{\var{(Q_{\mathrm{S}}')}}\con
1\quad \mbox{in  probability
and}\quad\frac{\sum_{k=1}^nE(G_{n,k}^4)}{\var^2({Q_{\mathrm{S}}'})}\con
0.
\end{align}
Define $\bfg_{k-1} = \sum_{i=1}^{k-1} \left(\v_i\v_i^T -
 p^{-1}\I_p\right).$ We have
\begin{align*}
\sum_{k=1}^n\sigma_{n,k}^2=&\sum_{k=1}^nE_{k-1}(G_{n,k}^2)\\
=&\sum_{k=1}^n4\{n(n-1)\}^{-2}p^2 \big(\v_k^T\bfg_{k-1}\v_k
\big)^2\\
=&\frac{8}{\{n(n-1)\}^2}\sum\limits_{k=1}^{n}\tr(\bfg_{k-1}^2).
\end{align*}
By noting that
\begin{align*}
\tr \bigg( \sum\limits_{k=1}^{n} \bfg_{k-1}^2 \bigg)
& =  \sum\limits_{k=1}^{n} \sum\limits_{i=1}^{k-1} \sum\limits_{j=1}^{k-1}\tr \left\{\left(\v_i\v_i^T - p^{-1}\I_p\right)\left(\v_j \v_j^T -  p^{-1}\I_p\right) \right\} \\
& =\frac{n(n-1)(p-1)}{2p} + \sum\limits_{i \neq j} 2
\left\{n-\max(i,j)\right\}\tr \left\{\left(\v_i \v_i^T -
p^{-1}\I_p\right)\left(\v_j \v_j^T -  p^{-1}\I_p\right) \right\},
\end{align*}
we can obtain
\begin{align*}
E\left(\sum\limits_{k=1}^{n} \sigma_{n,k}^2\right)= \frac{4(p-1)}{n(n-1)p},\
\var\left(\sum\limits_{k=1}^{n} \sigma_{n,k}^2\right) =
\frac{128(n-2)(p-1)}{3\{n(n-1)\}^3p^2(p+2)}.
\end{align*}
Clearly, $\sum_{k=1}^n\sigma^2_{n,k}/\var(Q_{\mathrm{S}}') \to 1$.

Finally, we verify that the second part of (\ref{lyacons}). Note that
\begin{align*}
\sum\limits_{k=1}^{n} E(G_{n,k}^4)
=&\frac{16p^4}{\{n(n-1)\}^4} \Bigg[ \frac{n(n-1)}{2} E\left\{\v_k^T\left(\v_i\v_i^T -  p^{-1}\I_p\right) \v_k\right\}^4 \\
& +n(n-1)(n-2) E\left\{\left(\v_k^T\left(\v_i \v_i^T -
p^{-1}\I_p\right)\v_k\right)^2\left(\v_k^T (\v_j \v_j^T -
p^{-1}\I_p) \v_k\right)^2\right\} \Bigg].
\end{align*}
Because
\begin{align*}
&E\left\{\v_k^T(\v_i\v_i^T- p^{-1}\I_p)\v_k\right\}^4= O(p^{-4}),\\
&E\left[\left\{\v_k^T(\v_i\v_i^T-
p^{-1}\I_p)\v_k\right\}^2\left\{\v_k^T(\v_j\v_j^T-p^{-1}\I_p)\v_k\right\}^2\right]=
O(p^{-4}),
\end{align*}
it is straightforward to see
${\sum_{k=1}^nE(G_{n,k}^4)}=o\{{\var^2({Q_{\mathrm{S}}'})}\}$. Here
we completes the proof of this lemma. \hfill$\Box$
\par
\vspace{0.5cm}

\setcounter{equation}{0} 
\noindent{\bf Appendix B: Proof of Theorems}

\vspace{0.2cm} \noindent{\bf Proof of Theorem 1} We decompose
$\U_{ij}$ as
\begin{align*}
\U_{ij}=U(\X_i-\X_j)=E(U(\X_i-\X_j)|\X_i)-E(U(\X_i-\X_j)|\X_j)+\bmw_{ij}
\end{align*}
Under $H_0$, $E(U(\X_i-\X_j)|\X_i)=\u_i$. Then,
$\U_{ij}=\u_i-\u_j+\bmw_{ij}$. Obviously, $E(\bmw_{ij})=0$,
$E(\u_i^T\bmw_{ij})=0$ and $E(\bmw_{ij}^T\bmw_{ik})=0$. And by Lemma
1, we have $E(\bmw_{ij}^T\bmw_{ij})=1-2\tau_F=o(1)$.
\begin{align*}
\tilde{Q}_S=&\frac{2p}{n(n-1)(n-2)(n-3)}\underset{i,j,k,l~are~not~equal}{\sum\sum\sum\sum}\U_{ij}^T\U_{kl}\U_{kj}^T\U_{il}-1\\
=&\left(\frac{4p}{n(n-1)}\underset{i\not=j}{\sum\sum}(\u_i^T\u_j)^2-1\right)-\frac{2p}{n(n-1)(n-2)}
\underset{i,j,k~are~not~equal}{\sum\sum\sum}\u_i^T\u_j\u_j^T\u_k\\
&+\frac{p}{n(n-1)(n-2)(n-3)}\underset{i,j,k,l~are~not~equal}{\sum\sum\sum\sum}\u_i^T\u_j\u_k^T\u_l\\
&+O(pn^{-4})\underset{i,j,k,l~are~not~equal}{\sum\sum\sum\sum}\Big(\u_i^T\u_j\u_i^T\bmw_{kl}+\u_i^T\u_j\u_k^T\bmw_{il}+\u_i^T\u_k\u_i^T\bmw_{kl}\\
&+\u_i^T\u_j\bmw_{kl}^T\bmw_{il}+\u_i^T\u_j\bmw_{ij}^T\bmw_{kl}+\u_i^T\bmw_{kl}\bmw_{kj}^T\bmw_{il}+\bmw_{ij}^T\bmw_{kl}\bmw_{kj}^T\bmw_{il}\Big)\\
\doteq & J_1+J_2+J_3+J_4
\end{align*}
According to Lemma 1 and 3, we have
\[J_1/\sigma_0 \cd N(0,1)\]
Thus, we only need to show the other parts are all $o_p(\sigma_0)$.
\begin{align*}
E(J_2^2)=&O(p^2n^{-2})E(\u_i^T\u_j\u_j^T\u_k\u_k^T\u_l\u_l^T\u_i)+O(p^2n^{-3})E(\u_i^T\u_j\u_j^T\u_k\u_k^T\u_j\u_j^T\u_i)\\
=&O(p^{-1}n^{-2})+O(p^{-1}n^{-3})=o(\sigma_0^2),\\
E(J_4^2)=&O(p^2n^{-4})E((\u_i^T\u_j\u_k^T\u_l)^2)=O(p^{-1}n^{-4})=o(\sigma_0^2).
\end{align*}
Finally, we only consider the first part in $J_4$. The proof of the other parts are similar.
\begin{align*}
E\Big(O(pn^{-4})&\underset{i,j,k,l~are~not~equal}{\sum\sum\sum\sum}\u_i^T\u_j\u_i^T\bmw_{kl}\Big)^2\\
=&O(p^2n^{-3})E(\u_i^T\u_j\u_i^T\bmw_{kl}\u_s^T\u_j\u_s^T\bmw_{kl})+O(p^2n^{-4})E((\u_i^T\u_j\u_i^T\bmw_{kl})^2)\\
=&O(p^{-1}n^{-3})E(\bmw_{kl}^T\bmw_{kl})+O(p^{-1}n^{-4})E(\bmw_{kl}^T\bmw_{kl})\\
=&o(p^{-1}n^{-3})+o(p^{-1}n^{-4})=o(\sigma_0^2).
\end{align*}
Here we complete the proof. \hfill$\Box$
\par

\vspace{0.2cm}
\noindent{\bf Proof of Theorem 2} Define
$\V_i=E(U(\X_i-\X_j)|\X_i)$. Similar to the arguments as Theorem 1,
we can show that
\begin{align*}
\tilde{Q}_S=&\frac{4p}{n(n-1)}\underset{i\not=j}{\sum\sum}(\V_i^T\V_j)^2-1+o_p(\sigma_1)
\end{align*}
Now, write
$\V_i=\{\BL_p^{1/2}\u_i\}/\{1+\u_i^T\D_{n,p}\u_i\}^{1/2}$, and then
\begin{align*}
E&(\V_i^T\V_j)^2
=\tr\left(\left[E\left\{\BL_p^{1/2}\u_i{\u_i}^T\BL_p^{1/2}(1+\u_i^T\D_{n,p}\u_i)^{-1}\right\}\right]^2\right)\\
=&\tr\left[\left\{E\left(\BL_p^{1/2}\u_i{\u_i}^T\BL_p^{1/2}\right)\right\}^2\right]
+\tr\left(\left[E\left\{C_i\BL_p^{1/2}\u_i{\u_i}^T\BL_p^{1/2}\left(\u_i^T\D_{n,p}\u_i\right)\right\}\right]^2\right),
\end{align*}
where $C_i$ is a bounded random variable between $-1$ and
$-(1+\u_i^T\D_{n,p}\u_i)^{-2}$. Obviously,
$\tr\left[\left\{E\left(\BL_p^{1/2}\u_i{\u_i}^T\BL_p^{1/2}\right)\right\}^2\right]=\tau_F^2p^{-2}\tr(\BL_p^2)=\tau_F^2p^{-2}(p+\tr(\D_{n,p}^2))$.
By the Cauchy inequality and Lemma 2,
\begin{align*}
\tr&\left(\left[E\left\{C_i\BL_p^{1/2}\u_i{\u_i}^T\BL_p^{1/2}\left(\u_i^T\D_{n,p}\u_i\right)\right\}\right]^2\right)\\
\le &C \tr\left[\left\{E\left(\BL_p^{1/2}\u_i{\u_i}^T\BL_p^{1/2}\right)^2 \right\}\right]E\left\{\left(\u_i^T\D_{n,p}\u_i\right)^2\right\} \\
\le & C p^{-4}\tr(\BL_p^2)\tr(\D_{n,p}^2)=
Cp^{-4}\{p+\tr(\D_{n,p}^2)\}\tr(\D_{n,p}^2)=o(p^{-1}n^{-1})
\end{align*}
by the condition $\tr(\D_{n,p}^2)=O(n^{-1}p)$. Consequently,
$E(Q_{\mathrm{S}}')=p\tr(\BL_p^2)-1+o(n^{-1}).$ Taking the same
procedure as $E\{(\V_i^T\V_j)^2\}$, we can obtain that
\begin{align*}
&E(\V_i^T\V_j)^4=\{3\tr^2(\BL_p^2)+6\tr(\BL_p^4)\}/\{p(p+2)(p+4)(p+6)\}[1+O\{p^{-2}\tr(\D_{n,p}^2)\}], \\
&E\left\{(\V_i^T\V_j)^2(\V_i^T\V_k)^2\right\}=\{\tr^2(\BL_p^2)+2\tr(\BL_p^4)\}/\{p^3(p+2)\}[1+O\{p^{-2}\tr(\D_{n,p}^2)\}].
\end{align*}
And then,
\begin{align*}
\var\bigg\{\frac{1}{n(n-1)}&\sum_{i\neq
j}(\V_i^T\V_j)^2\bigg\}=\left[\frac{4\tr^2(\BL_p^2)}{n(n-1)p^4}+\frac{8\{p\tr(\BL_p^4)-\tr^2(\BL_p^2)\}}{(n-1)p^4}\right]\{1+o(1)\}.
\end{align*}
Thus,
\begin{align*}
 E(\tilde{Q}_S)&={\tr(\D_{n,p}^2)}/p+o(n^{-1}),\\
 \var(\tilde{Q}_S)&=\left[\frac{4\tr^2(\BL_p^2)}{n(n-1)p^2}+\frac{8\{\tr(\BL_p^4)-p^{-1}\tr^2(\BL_p^2)\}}{(n-1)p^2}\right]\{1+ o(1)\}.
\end{align*}
It suffices to show that $T_n=\{n(n-1)\}^{-1}\sum_{i\neq
j}4p(\V_i^T\V_j)^2 $ is asymptotically normal. Obviously,
\begin{align*}
\var^2(T_n) \geq K \max
\bigg\{\frac{\{\tr(\BL_p^4)-p^{-1}\tr^2(\BL_p^2)\}\tr^2(\BL_p^2)}{n(n-1)^2p^4},
\frac{\tr^4(\BL_p^2)}{\{n(n-1)\}^2p^4} \bigg\}
\end{align*}
for sufficiently large $n$, where $K$ is some constant.

Then we also use the martingale central limit theorem (Hall and Hyde
1980) to prove the asymptotical normality. For this purpose, let
$\mathcal{F}_0=\{\varnothing,\bfo\}$,
$\mathcal{F}_k=\sigma\{\V_1,\ldots,\V_k\}, k = 1,\ldots,n$. Let
$E_k(\cdot)$ denote the conditional expectation of given
$\mathcal{F}_k$ and $E_0(\cdot)= E(\cdot)$. Write
$T_n-E(T_n)=\sum_{k=1}^nG_{n,k}$, where $G_{n,k}=(E_k-E_{k-1})T_n$.
Then for every $n$, $\{G_{n,k}\}_{k=1}^n$ is a martingale difference
sequence with respect to the $\sigma$-fields $\{\mathcal{F}_k,1\leq
k\leq n\}$. Let $\sigma_{n,k}^2=E_{k-1}(G_{n,k}^2)$. It suffices to
show that, as $n\con\infty$,
\begin{align}\label{lyacon}
\frac{\sum_{k=1}^n\sigma^2_{n,k}}{\var{(T_n)}}\con 1 \quad \mbox{in
probability
and}\quad\frac{\sum_{k=1}^nE(G_{n,k}^4)}{\var^2({T_n})}\con 0.
\end{align}
As $E(\sum_{k=1}^n\sigma^2_{n,k})=\var(T_n)$, to see the first part
of (\ref{lyacon}), we only show $\var(\sum_{k=1}^n\sigma^2_{n,k})=
o\{\var^2(T_n)\}$. Define $2E(\V_i\V_i^T)=\bfg_p$ and $\bfg_{k-1} =
\sum_{i=1}^{k-1} \left(2\V_i\V_i^T - \bfg_p\right).$ By the same
procedure as $E\{(\V_i^T\V_j)^2\}$,
\begin{align*}
\sigma_{n,k}^2 = & E_{k-1}(G_{n,k}^2) \\
= & \bigg[\frac{8p^2}{\{n(n-1)\}^2}\frac{\left\{ \tr(\bfg_{k-1}\BL_p )^2\tr^2(\BL_p)-\tr^2(\bfg_{k-1}\BL_p)\tr(\BL_p^2) \right\}}{\tr^4(\BL_p)} \\
& + \frac{16p^2}{n^2(n-1)}\frac{\left\{
\tr(\bfg_{k-1}\BL_p^3)\tr^2(\BL_p)-\tr(\bfg_{k-1}\BL_p)\tr^2(\BL_p^2)
\right\}}{\tr^5(\BL_p)} \\
&+\frac{8p^2}{n^2}\frac{\{\tr(\BL_p^4)-p^{-1}\tr^2(\BL_p^2)\}}{\tr^4(\BL_p)}\bigg]\left[1+o\{p^{-2}\tr(\D_{n,p}^2)\}\right].
\end{align*}
Then
\begin{align*}
\sum\limits_{k=1}^{n} \sigma_{n,k}^2 = (R_{1,n} + R_{2,n} + R_{3,n} +
R_{4,n} + R_{5,n} + C)\{1+o(1)\},
\end{align*}
where $ C $ is a constant, and { \begin{align*} R_{1,n}&=
\frac{32p^2}{\{n(n-1)\}^2}\frac{\tr^2(\BL_p^2)\sum^{n}_{k=1}(k-1)(\sum^{k-1}_{i=1}\V_i^T\BL_p\V_i)
}{\tr^5(\BL_p)},\\
R_{2,n}&= -
\frac{32p^2}{\{n(n-1)\}^2}\frac{\sum^{n}_{k=1}(k-1)(\sum^{k-1}_{i=1}\V_i^T\BL_p^3\V_i)
}{\tr^3(\BL_p)},\\
R_{3,n}&=
\frac{32p^2}{n^2(n-1)}\frac{(\sum^{n}_{k=1}\sum^{k-1}_{i=1}\V_i^T\BL_p^3\V_i)
}{\tr^3(\BL_p)},\\
R_{4,n}&= -
\frac{32p^2}{n^2(n-1)}\frac{\tr^2(\BL_p^2)(\sum^{n}_{k=1}\sum^{k-1}_{i=1}\V_i^T\BL_p\V_i)
}{\tr^5(\BL_p)},\\
R_{5,n}&=
\frac{32p^2}{\{n(n-1)\}^2}\frac{\sum^{n}_{k=1}\sum^{k-1}_{i=1}\sum^{k-1}_{j=1}(\V_i^T\BL_p
\V_j)^2}{\tr^2(\BL_p)}.
\end{align*}}
It suffices to show $\var(R_{i,n})= o\{\var^2(T_n)\}$ for
$i=1,\ldots,6$. Using
\begin{align*}
\var&\left\{\sum\limits_{k=1}^{n}(k-1)\left(\sum\limits_{i=1}^{k-1}\V_i^T\BL_p\V_i\right)\right\}\\
= & \left\{ \sum\limits_{i=1}^{n}\frac{(n-i)^2(n+i-1)^2}{4} \right\} \left[ E(\V_i^T\BL_p\V_i)^2 - \left\{ E(\V_i^T\BL_p\V_i) \right\}^2 \right] \\
= & \left\{ \sum\limits_{i=1}^{n}\frac{(n-i)^2(n+i-1)^2}{2} \right\}
\frac{\left\{\tr(\BL_p^4)-p^{-1}\tr^2(\BL_p^2)\right\}}{4p^2}\{1+o(1)\},
\end{align*}
we have
\begin{align*}
\frac{\var(R_{1,n})}{\var^2(T_n)} \leq
K\frac{\tr^2(\BL_p^2)}{\tr^4(\BL_p)}\con 0.
\end{align*}
By carrying out similar procedures we can show that $\var(R_{i,n})=
o\{\var^2(T_n)\}$ for $i=1,\ldots,6$, and hence complete the proof for
the first part of (\ref{lyacon}).

To show the second part of (\ref{lyacon}),
\begin{align*}
\sum\limits_{k=1}^{n} E(G_{n,k}^4) \leq & \frac{128p^4}{n^3}E\bigg\{2\V_k^T\bfg_p\V_k -\tr(\bfg_p^2)\bigg\}^4 \\
& + \frac{128p^4}{\{n(n-1)\}^4} \sum\limits_{k=1}^{n} E\bigg\{
2\V_k^T\bfg_{k-1}\V_k -\tr(\bfg_{k-1}\bfg_p) \bigg\}^4.
\end{align*}
By some algebra, we get
\begin{align*}
E\bigg\{2\V_k^T\bfg_p\V_k -\tr(\bfg_p^2)\bigg\}^4 \leq
K\frac{\tr(\BL_p^4)\left\{\tr(\BL_p^4)-p^{-1}\tr^2(\BL_p^2)\right\}}{\tr^{8}(\BL_p)},
\end{align*}
which leads to
\begin{align*}
\frac{\frac{128p^4}{n^3}E\bigg\{2\V_k^T\bfg_p\V_k
-\tr(\bfg_p^2)\bigg\}^4}{\var^2(T_n)} \leq
K\frac{\tr(\BL_p^4)}{\tr^2(\BL_p^2)}.
\end{align*}
By the Cauchy inequality, $\tr(\D_{n,p}^4)\le \tr^2(\D_{n,p}^2)$ and
$\tr^2(\D_{n,p}^3)\le \tr(\D_{n,p}^4)\tr(\D_{n,p}^2)$, so
$\tr(\BL_p^4)=o(p^2)=o(\tr^2(\BL_p^2))$ by the condition
$\tr(\D_{n,p}^2)=O(n^{-1}p)$. Thus,
$\frac{128p^4}{n^3}E\bigg\{2\V_k^T\bfg_p\V_k
-\tr(\bfg_p^2)\bigg\}^4=o(\var^2(T_n))$. Similarly, we can get
\begin{align*}
\frac{128p^4}{\{n(n-1)\}^4} \sum\limits_{k=1}^{n} E\bigg\{
2\V_k^T\bfg_{k-1}\V_k -\tr(\bfg_{k-1}\bfg_p) \bigg\}^4 =
o\{\var^2(T_n)\}.
\end{align*}
Here we can complete the proof for the second part of
(\ref{lyacon}). \hfill$\Box$

\par
\vspace{0.2cm}
\noindent{\bf Proof of Theorem 3}
Under $H_0$, similar to $\tilde{Q}_S$, we decompose $\tilde{Q}_K$ as follow,
\begin{align*}
\tilde{Q}_K=&\frac{p}{n(n-1)(n-2)(n-3)}\underset{i,j,k,l~are~not~equal}{\sum\sum\sum\sum} (\U_{ij}^T\U_{kl})^2-1\\
=&\frac{p}{n(n-1)(n-2)(n-3)}\underset{i,j,k,l~are~not~equal}{\sum\sum\sum\sum} ((\u_i-\u_j+\bmw_{ij})^T(\u_k-\u_l+\bmw_{kl}))^2-1\\
=&\frac{4p}{n(n-1)}\underset{i\not=j}{\sum\sum}(\u_i^T\u_j)^2-4\tau_F^2-\frac{2p}{n(n-1)(n-2)}\underset{i,j,k~are~not~equal}{\sum\sum\sum}\u_i^T\u_j\u_j^T\u_k\\
&+\frac{p}{n(n-1)(n-2)(n-3)}\underset{i,j,k,l~are~not~equal}{\sum\sum\sum\sum}\u_i^T\u_j\u_k^T\u_l\\
&+O(pn^{-3})\underset{i,j,k~are~not~equal}{\sum\sum\sum}\u_i^T\u_j\u_i^T\bmw_{jk}+O(pn^{-4})\underset{i,j,k,l~are~not~equal}{\sum\sum\sum\sum}\u_i^T\u_k\u_j^T\bmw_{kl}\\
&+O(pn^{-3})\underset{i,j,k,l~are~not~equal}{\sum\sum\sum\sum}\u_i^T\u_k\bmw_{ij}^T\bmw_{kl}\\
&+O(pn^{-3})\underset{i,j,k~are~not~equal}{\sum\sum\sum}((\u_i^T\bmw_{jk})^2-p^{-1}(1-2\tau_F))\\
&+O(pn^{-4})\underset{i,j,k,l~are~not~equal}{\sum\sum\sum\sum}((\bmw_{ij}^T\bmw_{kl})^2-(1-2\tau_F)^2)
\end{align*}
According to the proof of Theorem 1, we only need to show the last two parts are $o_p(\sigma_0^2)$.
\begin{align*}
E&\Big(O(pn^{-3})\underset{i,j,k~are~not~equal}{\sum\sum\sum}\left((\u_i^T\bmw_{jk})^2-p^{-1}(1-2\tau_F)\right)\Big)^2\\
=&O(p^2n^{-3})E(\left((\u_i^T\bmw_{jk})^2-p^{-1}(1-2\tau_F)\right)^2)\\
&+O(p^2n^{-2})E\left(\left((\u_i^T\bmw_{jk})^2-p^{-1}(1-2\tau_F)\right)\left((\u_l^T\bmw_{jk})^2-p^{-1}(1-2\tau_F)\right)\right)\\
=&O(p^2n^{-3})\left(E((\u_i^T\bmw_{jk})^4)-p^{-2}(1-2\tau_F)^2\right)\\
+&O(p^2n^{-2})\left(E((\u_i^T\bmw_{jk})^2(\u_l^T\bmw_{jk})^2)-p^{-2}(1-2\tau_F)^2\right)\\
=&o(n^{-3})+o(n^{-2})=o(\sigma_0^2),\\
E&\left(O(pn^{-4})\underset{i,j,k,l~are~not~equal}{\sum\sum\sum\sum}((\bmw_{ij}^T\bmw_{kl})^2-(1-2\tau_F)^2)\right)^2\\
=&O(p^2n^{-4})E((\bmw_{ij}^T\bmw_{kl})^4-(1-2\tau_F)^2)+O(p^2n^{-2})E((\bmw_{ij}^T\bmw_{kl})^2(\bmw_{is}^T\bmw_{kt})^2-(1-2\tau_F)^2)\\
=&o(n^{-2})=o(\sigma_0^2).
\end{align*}
Thus, we proof result (i). Similarly, we can also proof the result (ii) under $H_1$.\hfill$\Box$
\par
\vspace{0.5cm}
\noindent{\bf Appendix C: Proof of Corollaries}

\vspace{0.2cm}
\noindent{\bf Proof of Corollary 1}
From Theorems 1-2,
\begin{align*}
\lim\inf_n
pr\left(\frac{\tilde{Q}_S-p\delta_{n,p}}{{\sigma}_0}>z_{\alpha}\right)\geq
1-\lim\sup_n\Phi\left\{\frac{{\sigma}_0z_{\alpha}-p^{-1}\tr(\D_{n,p}^2)}{{\sigma}_1}\right\}.
\end{align*}
Obviously, ${\sigma}_0/{\sigma}_1=O(1)$ due to
$\tr(\BL_p^4)-p^{-1}\tr^2(\BL_p^2)\geq 0$. Denote
\begin{align*}
\gamma_{1n}&=\frac{8\left\{\tr(\BL_p^4)-p^{-1}\tr^2(\BL_p^2)\right\}}{p^2},\\
\gamma_{2n}&=\frac{8\left\{\tr(\BL_p^4)\tr^2(\BL_p)+\tr^3(\BL_p^2)-2\tr(\BL_p)\tr(\BL_p^2)\tr(\BL_p^3)\right\}}{\tr^2(\BL_p^2)p^2}.
\end{align*}
Firstly, consider the case $p/\tr(\D_{n,p}^2)=o(1)$. The condition
$n\tr(\D_{n,p}^2)/p\con\infty$ leads to
\begin{align*}
\frac{{\sigma}_1^2}{p^{-2}\tr^2(\D_{n,p}^2)}&=O\left\{\frac{p^2}{n^2\tr^2(\D_{n,p}^2)}\right\}+O\left\{\frac{\tr(\BL_p^4)}{n\tr^2(\D_{n,p}^2)}\right\}\\
&=O\left\{\frac{\tr^2(\D_{n,p}^2)}{n\tr^2(\D_{n,p}^2)}\right\}+o(1)\con
0,
\end{align*}
which implies the assertion of Corollary 1. For the case
$p/\tr(\D_{n,p}^2)=O(1)$, it can be seen that
$\gamma_{2n}/\gamma_{1n}=O(1)$. By Theorem 4-(i) in Chen et al.
(2010), we have $\gamma_{2n}/\{np^{-2}\tr^2(\D_{n,p}^2)\}\con 0$
from which the corollary follows immediately. \hfill$\Box$
\par
\vspace{0.2cm}
\noindent{\bf Proof of Corollary 2}
By Theorem 1 in Chen et al. (2010),
\begin{align*}
 \frac{C_n-\tr(\D_{n,p}^2)/p}{\sqrt{4{n^{-2}}+\gamma_{2n}n^{-1}}}\con N(0,1)
\end{align*}
in distribution, where $C_n$ is the test statistic proposed by Chen
et al.  (2010). Thus, the power function of $C_n$ is
\begin{align*}
\beta_{C_n}=\Phi\left(-\frac{2n^{-1}}{\sqrt{4{n^{-2}}+\gamma_{2n}n^{-1}}}z_{\alpha}+\frac{\tr(\D_{n,p}^2)/p}{\sqrt{4{n^{-2}}+\gamma_{2n}n^{-1}}}\right).
\end{align*}
According to Theorem 1 and 2, the power function of $\tilde{Q}_S$ is
\begin{align*}
\beta_{\tilde{Q}_S}=\Phi\left(-\frac{\sigma_0}{\sigma_1}z_{\alpha}+\frac{\tr(\D_{n,p}^2)/p}{\sigma_1}\right).
\end{align*}
Obviously, $\sigma_0=2n^{-1}(1+o(1))$ as $p \to \infty$. Then, the
asymptotic relative efficiency of $\tilde{Q}_S$ with respect to
$C_n$ is one in this case. \hfill$\Box$
\par
\vspace{0.2cm}
\noindent{\bf Proof of Corollary 3}
According to the proof of Theorem 3 (ii),
$\tilde{Q}_K=\tilde{Q}_S+o_p(\sigma_1)$. Thus, by Corollaries 1 and
2, we can easily obtain the results.\hfill$\Box$

 \vspace{1cm}

\noindent{\bf {References}}:

\begin{description}

\item  {\sc Bai, Z.} and {\sc Saranadasa, H.} (1996),  Effect of High Dimension: by an
Example of a Two Sample Problem,  {\it Statist. Sinica, } { 6},
311--329.

\item {\sc Bai, Z., Jiang, D., Yao, J.} and {\sc Zheng, S.} (2009)
\newblock Corrections to LRT on large-dimensional covariance matrix by
RMT. {\it Ann. Statist.} {\bf 37},3822--3840.

\item
\textsc{Chen, S.~X.} and \textsc{Qin, Y.~L.} (2010).
\newblock A two-sample test for high-dimensional data with applications to
gene-set testing.
\newblock \textit{Ann. Statist.} \textbf{38}, 808--835.

\item
\textsc{Chen, S.~X.}, \textsc{Zhang, L.~X.} and \textsc{Zhong,
P.~S.} (2010).
\newblock Tests for high-dimensional covariance matrices.
\newblock \textit{J. Am. Statist. Assoc.} \textbf{105}, 801--815.

\item
 {\sc
Hall, P. G.} and {\sc Hyde, C. C.} (1980). {\em Martingale central
limit theory and its applications}. \newblock Academic Press, New
York.

\item
\textsc{Hallin, M.} and \textsc{Paindaveine, D.} (2006).
\newblock Semiparametrically efficient rank-based inference for shape.
I: Optimal rank-based tests for sphericity.
\newblock \textit{Ann. Statist.} \textbf{34}, 2707--2756.

\item
\textsc{Ledoit, O.} and \textsc{Wolf, M.} (2002).
\newblock Some hypothesis tests for the
covariance matrix when the dimension is large compared to the sample size.
\newblock \textit{Ann. Statist.} \textbf{30}, 1081--1102.

\item
\textsc{John, S.} (1971).
\newblock Some optimal multivariate tests.
\newblock \textit{Biometrika} \textbf{59}, 123--127.

\item
\textsc{John, S.} (1972).
\newblock The distribution of a statistic used for testing sphericity of normal distributions.
\newblock \textit{Biometrika} \textbf{59}, 169--173.

\item
\textsc{Marden, J.} and \textsc{Gao, Y.} (2002).
\newblock Rank-based procedures for structural hypotheses on covariance matrices.
\newblock \textit{Sankhy$\bar{a}$ Ser. A} \textbf{64}, 653--677.


\item
\textsc{Mauchly, J.~W.} (1940).
\newblock Significance test for sphericity of a normal n-variate distribution.
\newblock \textit{Ann. Math. Statist.} \textbf{11}, 204--209.

\item
\textsc{M\"{o}tt\"{o}nen  J.} and \textsc{Oja, H.} (1995).
\newblock Multivariate spatial sign and rank methods.
\newblock \textit{J. Nonparametr. Statist} \textbf{5}, 201--213.

\item
\textsc{Muirhead, R.~J.} and \textsc{Waternaux, C.~M.} (1980).
\newblock Asymptotic distributions in canonical correlation analysis and other
multivariate procedures for nonnormal populations.
\newblock \textit{Biometrika} \textbf{67}, 31--43.

\item
\textsc{Oja, H.} (2010).
\newblock \textit{{Multivariate Nonparametric Methods with R}}.
\newblock Springer, New York.

\item
\textsc{Onatski, A.}, \textsc{Moreira, M.~J.} and \textsc{Hallin,
M.} (2013).
\newblock Asymptotic power of sphericity tests for high-dimendional data.
\newblock \textit{Ann. Statist.} \textbf{41}, 1204--1231.

\item
\textsc{Sirki\"{a}, S.}, \textsc{Taskinen, S.}, \textsc{Oja, H.} and
\textsc{Tyler, D.~E.} (2009).
\newblock Tests and estimates of shape based on spatial signs and ranks.
\newblock \textit{J. Nonparametr. Statist.} \textbf{21}, 155--176.

\item {\sc Srivastava, M. S., Kollo, T.} and {\sc von Rosen, D.} (2011),  Some tests for the covariance matrix with fewer observations than
the dimension under non-normality,  {\it J. Multivar. Anal.}, {102},
1090--1103.

\item
\textsc{Tyler, D.~E.} (1987). Statistical analysis for the angular
central Gaussian distribution on the sphere.
\newblock \textit{Biometrika} \textbf{74}, 579--589.

\item {\sc Wang, L., Peng, B.} and {\sc Li, R.} (2015).  A high-dimensional
nonparametric multivariate test for mean vector,  {\it J. Am.
Statist. Assoc.}, To appear.

\item {\sc Zou, C., Peng, L, Feng, L.} and {\sc Wang, Z.}
(2014),  \newblock Multivariate-sign-based high-himensional tests
for sphericity,  {\it Biometrika}, { 101}, 229--236.

\end{description}

\end{document}